\documentclass[aoas]{imsart}

\RequirePackage{amsthm,amsmath,amsfonts,amssymb}
\RequirePackage{natbib}
\usepackage{natbib}
\usepackage{amsfonts,amsmath,amssymb}
\usepackage[caption=false]{subfig}
\usepackage{graphicx}
\usepackage{bm}
\usepackage{booktabs}
\usepackage{enumitem}
\usepackage{mathtools}
\usepackage{hyperref}
\def\bSig\mathbf{\Sigma}

\newcommand{\indep}{\rotatebox[origin=c]{90}{$\models$}}
\DeclareMathOperator*{\argmax}{arg\,max}

\startlocaldefs

\endlocaldefs

\begin{document}

\begin{frontmatter}
\title{Identifying optimally cost-effective dynamic treatment regimes with a Q-learning approach}
\runtitle{Optimally cost-effective dynamic treatment regimes}

\begin{aug}
\author[A]{\fnms{Nicholas} \snm{Illenberger}\ead[label=e1,mark]{nillen@pennmedicine.upenn.edu}},
\author[B]{\fnms{Andrew J.} \snm{Spieker}\ead[label=e2]{andrew.spieker@vumc.org}}
\and
\author[A]{\fnms{Nandita} \snm{Mitra}\ead[label=e3,mark]{nmitra@pennmedicine.upenn.edu}}
\address[A]{University of Pennsylvania, \printead{e1,e3}}

\address[B]{Vanderbilt University Medical Center, \printead{e2}}
\end{aug}

\begin{abstract}
Health policy decisions regarding patient treatment strategies require consideration of both treatment effectiveness and cost. Optimizing treatment rules with respect to effectiveness may result in prohibitively expensive strategies; on the other hand, optimizing with respect to costs may result in poor patient outcomes. We propose a two-step approach for identifying an optimally cost-effective and interpretable dynamic treatment regime. First, we develop a combined Q-learning and policy-search approach to estimate an optimal list-based regime under a constraint on expected treatment costs. Second, we propose an iterative procedure to select an optimally cost-effective regime from a set of candidate regimes corresponding to different cost constraints. Our approach can estimate optimal regimes in the presence of time-varying confounding, censoring, and correlated outcomes. Through simulation studies, we illustrate the validity of estimated treatment regimes and examine operating characteristics under flexible modeling approaches. We also apply our methodology to evaluate optimally cost-effective treatment strategies for assigning adjuvant therapies to endometrial cancer patients.
\end{abstract}

\end{frontmatter}

\section{Introduction}

Due to rising healthcare costs, there has been growing interest in improved methods for cost-effectiveness analyses. Cost-effectiveness research is concerned with identifying policies that can balance treatment effectiveness and overall costs. Adjuvant radiation or chemotherapy is occasionally recommended to endometrial cancer patients undergoing hysterectomy because these therapies can reduce the risk of locoregional recurrence \citep{van2021adjuvant}. However, these therapies are also associated with higher toxicity rates and greater treatment costs \citep{randall2006randomized}. Identifying treatment sequences that are both effective and cost-efficient is complicated by heterogeneous treatment responses across individuals. Patients at high-risk of cancer recurrence are likely to benefit more from adjuvant therapies than are patients at low-risk of recurrence \citep{van2021adjuvant}. These complications underscore the need for treatment regime methodology that can optimally allocate limited resources towards patients that are most likely to benefit.

Traditionally, regression approaches have been used to address treatment response heterogeneity and its effects on cost-effectiveness. \cite*{willan2004regression} show that if cost and effectiveness metrics are linear functions of treatment status and covariates, then the incremental cost-effectiveness ratio (ICER), one measure of cost-effectiveness, can be represented as a function of the parameters in a least-squares regression and standard inferential procedures can be used to assess cost-effectiveness within subgroups. Alternatively, \cite*{nixon2005methods} develop Bayesian Markov Chain Monte Carlo methods for inference that are more robust to skewed costs. While these methods can describe cost-effectiveness within subgroups, neither allows for estimation of an optimal treatment strategy. \cite{lakkaraju2017learning} and \cite{xu2020estimating} develop methodology to learn optimal cost-effective individualized treatment rules in settings with point-exposures. Both propose regimes that maximize treatment effectiveness while placing a penalty on overall treatment cost. \cite{laber2018identifying} provides policy search methodology to identify a maximally effective treatment regime from among regimes with a pre-determined threshold on overall cost (defined by number of safety events). In contrast with other proposed methods, this approach can accomodate time-varying treatments. However, the decision rules within these regimes take the form of linear decision boundaries and may be difficult to interpret. Recent work by \cite{zhang2018interpretable} argues that treatment rules based on decision lists provide both flexibility and interpretability. 

\par In this paper, we propose an efficient algorithm for identifying optimal list-based decision rules under fixed cost constraints. We use Q-learning with policy search methodology to maximize treatment effectiveness within the class of list-based regimes with cost constraints. We modify the algorithm proposed by \cite{zhang2018interpretable} to efficiently estimate our list-based decision rules. Within our data example, we propose an iterative cost-effectiveness analysis that uses ICER to select an optimally cost-effective regime for the adjuvant treatment of endometrial cancer from a set of candidate regimes corresponding to different cost-constraints.

The remainder of this manuscript is organized as follows. In Section \ref{sec:methods} we introduce methodology for estimating optimal list-based treatment regimes under a constraint on overall expected treatment costs. We explore the operating characteristics of our proposed regime identification approach through simulations in Section \ref{sec:sim}. In Section \ref{sec:data_example} we perform a cost-effectiveness analysis to identify an optimal cost-effective treatment strategy for assigning endometrial cancer patients to adjuvant radiation and chemotherapy. We describe additional considerations and avenues for future developments in Section \ref{sec:discussion}. 

\section{Methods}\label{sec:methods}
\subsection{Set-up and notation}
\par Assume that treatment decisions are made at the beginning of $K$ distinct intervals indexed by $k = 1,...,K$. For each individual $i = 1,...,n$ in interval $k$, we observe $\left(W_{ik}, A_{ik}, C_{ik}, Z_{ik}, Y_{ik} \right)$. Here, $W_k$ denotes confounding variables collected at the beginning of interval $k$, $A_k$ denotes treatment status, $C_k$ is a censoring indicator, and $Z_k$ and $Y_k$ are clinical effectiveness and cost outcomes collected at the end of interval $k$.  Within each interval, variables are observed in the order: $W_k \longrightarrow A_k \longrightarrow C_k \longrightarrow (Z_k, Y_k)$. We use overbar notation to denote covariate history, e.g. $\overline{A}_k = (A_1, A_2, ...,A_{k-1}, A_k)$, and underbar notation to denote future values of a covariate, e.g. $\underline{A}_k = (A_{k+1}, A_{k+2}, ..., A_K)$. Let $H_k = (\overline{W}_k, \overline{A}_{k-1}, \overline{Z}_{k-1}, \overline{Y}_{k-1})$ denote a patient's covariate and treatment history at prior to decision $k$. 

\par A dynamic treatment regime is a set of decision rules $d = \{d_1,...,d_K\}$ where $d_k$ is a mapping $d_k : \mathcal{H}_k \longrightarrow \mathcal{A}_k$ from the space of all possible covariate histories into that of treatment decisions. Under a potential outcomes framework, if $\mathcal{D}$ is a class of treatment regimes, then the optimal dynamic treatment regime within this class, $d^{\text{opt}}$, satisfies the condition $\mathbb{E}[Z^{d^{\text{opt}}}_K] \ge \mathbb{E}[Z^{d}_K]$ for all $d \in \mathcal{D}$. Because a common goal in health policy is making treatment decisions within resource-limited settings, we restrict consideration to the class of list-based treatment rules with $\mathbb{E}[Y^d_K] < \tau$ for a predetermined cost-constraint, $\tau$. Note that $\tau$ constrains the expected cost for the entire population. Individual costs under this regime may exceed this constraint. This restricted search sacrifices regime flexibility in favor of interpretability. Following \cite{zhang2018interpretable}, a list-based regime is a regime for which each treatment rule $d_k$ for $k = 1,...,K$ consists of a series of if-else statements: 
\begin{itemize}[noitemsep, topsep = 1pt, labelwidth = 1pt]
    \item[{ }] \textbf{If} $H_{k} \in R_{k1}$ then $A_k = a_{k1}$;
    \item[{ }] \textbf{else if} $H_{k} \in R_{k2}$ then $A_k = a_{k2}$;
    \item[{ }] $\vdots$ 
    \item[{ }] \textbf{else if} $H_{k} \in R_{k L_k}$ then $A_k = a_{k L_k}$
\end{itemize} 
Here, $L_k$ is the preset maximal list length for a decision at interval $k$, and $R_{kl}$ is a subset of $\mathcal{H}_k$ for $l = 1,...,L_k$. For simplicity and to aid interpretability, we restrict $R_{kl}$ to clauses involving thresholding of a single covariate (e.g. $R_{kl} = \{h_k \in \mathcal{H}_k: h_{kj} \le \theta\}$ for $1 \le j \le \text{dim}(H_k), \ \theta \in \mathbb{R}$).
\par Estimation of the optimal cost-restricted regime involves identifying optimal values for $\{(R_{kl}, a_{kl}): l = 1,...,L_k\}_{k=1}^K$. We employ an integrated Q-learning and policy search approach to define and estimate these values. Define the $K^{th}$ stage Q-functions for $Z$ and $Y$:
\begin{align*}
    Q^Z_K(a_K, h_K) = \mathbb{E}[Z_K|A_K = a_K, H_K = h_K, C_K = 0] \\
    Q^Y_K(a_K, h_K) = \mathbb{E}[Y_K|A_K = a_K, H_K = h_K, C_K = 0]
\end{align*}
For a treatment regime $d$, we may recursively define Q-functions for intervals $k=K-1,...,1$:
\begin{align*}
    Q^Z_k(a_k, h_k; d) = \mathbb{E}[Z_k + Q^Z_{k+1}(d_{k+1}(H_{k+1}), H_{k+1})|A_k = a_k, H_k = h_k, C_k = 0], \\
    Q^Y_k(a_k, h_k; d) = \mathbb{E}[Y_k + Q^Y_{k+1}(d_{k+1}(H_{k+1}), H_{k+1})|A_k = a_k, H_k = h_k, C_k = 0].
\end{align*}
By the principles of dynamic programming, the optimal treatment regime can be identified by optimizing over the Q-functions at each individual decision point \citep{bellman1966dynamic}. In the policy search context, the process of finding an optimal regime within a pre-specified class reduces to finding the optimal decision rule within this class at each decision point. Because the $k^{\text{th}}$ stage Q-function for cost denotes the expected cost accrued in the $k^{\text{th}}$ interval and in future intervals under a pre-specified regime, decompose the overall cost constraint $\tau$ into $K$ components $\tau_1,...,\tau_K$ representing the interval specific cost constraints where $\sum_{k=1}^K \tau_k = \tau$. The optimal choices of $(R_K, a_K)$ within the class of list-based and cost constrained decision rules are given by:
\begin{align*}
     \{R_K^\text{opt}, a_K^{\text{opt}}\} =\ &\underset{R_K, a_K}{\argmax}\ \mathbb{E}\left[Q^Z_K\left(d_K(H_k; R_K, a_K), H_K\right)\right] \\
     &\text{subject to} \ \mathbb{E}\left[Q^Y_K(d(H_K; R_K, a_K), H_K)\right] < \tau_K.
\end{align*}
For decision intervals $k = K-1, ..., 1$ the optimal choices $(R_k, a_k)$ are given by:
\begin{align*}
     \{R^\text{opt}_k, a^\text{opt}_k\} =\ &\underset{R_k, a_k}{\argmax}\ \mathbb{E}\left[Q^Z_k\left(d_K(h_k; R_k, a_k), h_k; \underline{R}_k^\text{opt}, \underline{a}_k^\text{opt} \right)\right] \\
     &\text{subject to} \ \mathbb{E}\left[ Q^Y_k(d(H_k; R_k, a_k), H_k; ; \underline{R}_k^\text{opt}, \underline{a}_k^\text{opt})\right] < \sum_{j=k}^K \tau_k.
\end{align*}

To connect the distribution of the observed covariates to that of the potential outcomes we invoke the following identification assumptions:
\begin{itemize}
    \item[(A1)] (Positivity) If $P(H_k = h_k) > 0$, then $P(A_k = a_k|H_k = h_k) > 0$ for all $a_k$
    \item[(A2)] (Consistency) $Z_k = Z_k^{\overline{A}_k}$ and $Y_k = Y_k^{\overline{A}_k}$ for $k = 1,...,K$
    \item[(A3)](Sequentially ignorable treatment assignment) $(Z_k^{\overline{a}_k}, Y_k^{\overline{a}_k}, L_{k+1}^{\overline{a}_k}) \indep A_k | H_k$ for $k = 1,...,K$
    \item[(A4)](Sequentially ignorable censoring) $(Z_k^{\overline{a}_k}, Y_k^{\overline{a}_k}, L_{k+1}^{\overline{a}_k}) \indep C_k | A_k, H_k$ for $k = 1,...,K$
\end{itemize}
Under these assumptions, \cite{schulte2014q} and \cite{laber2018identifying} show: 
\begin{align*}
    Q^Z_K(a_K, h_K) &= \mathbb{E}[Z_K^{\overline{a}_{K-1}, a_K, C_K = 0}|H_K^{\overline{a}_{K-1}, a_K} = h_K] \\
    Q^Y_K(a_K, h_K) &= \mathbb{E}[Y_K^{a_{K-1}, a_{K}, C_K = 0}|H^{\overline{a}_{K-1}, a_K}_K = h_K]
\end{align*} 
and, for $k=1,...,k-1$:
\begin{align*}
    Q^Z_k(a_k, h_K;d) &= \mathbb{E}\bigg[Z_k^{\overline{a}_{k-1}, a_k, c_k = 0} + \sum_{j=k+1}^K Z_j^{\overline{a}_{k-1},a_k, \underline{c}_k = 0, \underline{d}_{k+1}}\bigg|H_k^{\overline{a}_{k-1}, a_k} = h_k\bigg] \\
    Q^Y_k(a_k, h_K;d) &= \mathbb{E}\bigg[Y_k^{\overline{a}_{k-1}, a_k, c_k = 0} + \sum_{j=k+1}^K Y_j^{\overline{a}_{k-1},a_k, \underline{c}_k = 0, \underline{d}_{k+1}}\bigg|H_k^{\overline{a}_{k-1}, a_k} = h_k\bigg]
\end{align*}
It follows that identifying optimal choices of $\{(R_{kl}, a_{kl}): l = 1,...,L_k\}_{k=1}^K$ maximizes the potential treatment effectiveness under a constraint on potential cost. 
\par In the next subsection, we propose a modification of the algorithm developed by \cite{zhang2018interpretable} for identifying globally optimal list-based regimes. Our extension allows for regimes to be fit under a preset cost constraints. 

\subsection{Estimating the Optimal Decision Rules}
To estimate the optimal list-based and cost constrained regime, we use a backwards recursive procedure. The optimal decision rule for the final interval is estimated first, and earlier timepoints are estimated assuming optimal decisions are made at all future timepoints. For each decision point $k$, we sequentially estimate the pairs $(R_{kl}, a_{kl})$ for clauses $l=1,...,L_k$. 
\par We illustrate estimation of the final decision rule, $d_K$, before describing estimation for earlier intervals. For a unit entering the final interval with covariate history $h_K$, define the unconstrained optimal rule $\widetilde{d}_K(h_K) = \underset{a_K}{\argmax} \ Q^Z_K \left(a_K, h_K \right)$. We want to approximate the optimal rule within the class of interpretable list-based and cost constrained treatment rules. To estimate the first clause of the constrained rule, define the intermediate list-based decision rule:
\begin{itemize}
    \item[] \textbf{If} $h_{K} \in R_{K1}$ then $A_K = a_{K1}$;
    \item[] \textbf{else if} $H_{K} \in \mathcal{H}_K$ then $A_K = \widetilde{d}_K(h_K)$
\end{itemize}
Given estimates of the $K^{th}$ stage Q-functions, the estimated mean effectiveness and cost measures under this regime are given by:
\begin{align*}
     \Psi^Z_{K1}(R_{K1}, a_{K1}) &= \frac{1}{n}\sum_{i=1}^n \left[ \mathbb{I}(h_{Ki} \in R_{K1}) \widehat{Q}^Z_K(a_{K1}, h_{Ki}) + \mathbb{I}(h_{Ki} \not\in R_{K1})\widehat{Q}^Z_K(\widetilde{d}_K(h_{Ki}), h_{Ki}) \right] \\
     \Psi^Y_{K1}(R_{K1}, a_{K1}) &=\frac{1}{n}\sum_{i=1}^n \left[ \mathbb{I}(h_{Ki} \in R_{K1}) \widehat{Q}^Y_K(a_{K1}, h_{Ki}) + \mathbb{I}(h_{Ki} \not\in R_{K1})\widehat{Q}^Y_K(\widetilde{d}_K(h_{Ki}), h_{Ki}) \right].
\end{align*}
We search for the values of $(R_{K1}, a_{K1})$ which maximize $\Psi^Z_{K1}$ while ensuring that the expected cost, $\Psi^Y_{K1}$, is less than or equal to the cost constraint, $\tau_K$. As \cite{zhang2018interpretable} point out, maximizing the objective function, $\Psi^Z_{K1}$, is equivalent to minimizing the difference between this intermediate rule and the optimal rule. Imposing the constraint $\Psi^Y_{K1} < \tau_K$ ensures that we are minimizing within the class of cost constrained rules. Different choices of $(R_{K1}, a_{K1})$ may lead to equivalent expected effectiveness and cost estimates. In these cases, we opt to reward decision regions that assign treatment to a greater number of patients by adding a complexity term, $\eta$, to our objective function: $\Psi^Z_{K1}(R_{K1}, a_{K1}) + \eta \left\{ \sum_{i=1}^n \mathbb{I}(h_{Ki} \in R_{K1}) \right\}$. Because $\eta$ may reward ``larger" regions at the expense of mean effectiveness, cross-validation can be used to select this parameter and ensure maximal effectiveness. The optimal choices of $(R_{K1}, a_{K1})$ for the intermediate rule are given by:
\begin{align*}
    (\widehat{R}_{K1}, \widehat{a}_{K1}) = &\underset{R_{K1},\ a_{K1}}{\argmax} \ \Psi^Z_{K1}(R_{K1}, a_{K1}) + \eta \left\{ \sum_{i=1}^n \mathbb{I}(h_{Ki} \in R_{K1}) \right\} \nonumber \\
    &\text{subject to} \ \Psi^Y_{K1}(R_{K1}, a_{K1}) < \tau_K.
\end{align*}
This procedure can be generalized to estimate optimal regions and treatment choices for each of the $L_K$ clauses in the list-based rule. The algorithm can be summarized as follows:
\begin{itemize}
    \item[Step 1.] Let $l = 1$.
    \item[Step 2.] Define $\widehat{G}_{Kl} = \mathcal{H}_K \setminus \left(\bigcup_{s<l} \widehat{R}_{Ks} \right)$. If $l = L_K$, force $R_{Kl} = \mathcal{H}_K$. The estimated mean of $X = Z$ or $Y$ under the $l^{\text{th}}$ intermediate rule is defined as:
     \begin{align} \label{eqn:dtr-algorithm}
        \Psi^X_{Kl}(R_{Kl}, a_{Kl}) = \frac{1}{n}\sum_{i=1}^n \bigg\{&\mathbb{I}(h_{Ki} \in R_{Kl}, h_{Ki} \in \widehat{G}_{Kl})\widehat{Q}^X_K(a_{Kl}, h_{Ki}) +  \\ 
            &\mathbb{I}(h_{Ki} \not\in R_{Kl}, h_{Ki} \in \widehat{G}_{Kl})\widehat{Q}^X_K(\widetilde{d}_K(h_{Ki}), h_{Ki}) + \nonumber \\
           &\sum_{j=1}^{l-1} \mathbb{I}(h_{Ki} \in \widehat{R}_{Kj}, h_{Ki}\not\in \widehat{G}_{Kj})\widehat{Q}^X_K(a_{Kj}, h_{Ki})\bigg\} \nonumber.
    \end{align}
    \item[Step 3.] Define $(\widehat{R}_{Kl}, \widehat{a}_{Kl})$ as:
    \begin{align*}
        (\widehat{R}_{Kl}, \widehat{a}_{Kl}) = &\argmax_{R_{Kl}, a_{Kl}} \  \Psi^Z_{Kl}(R_{Kl}, a_{Kl}) + \frac{1}{n}\sum_{i=1}^n\mathbb{I}(h_{Ki} \in R_{Kl}, h_{Ki} \in \widehat{G}_{Kl}); \nonumber \\
    &\text{ subject to } \Psi^Y_{Kl}(R_{Kl}, a_{Kl}) < \tau_K.
    \end{align*}
    If $l < L_K$ set $l = l + 1$ and return to Step 2, otherwise stop.
\end{itemize}
\par The algorithm above can be used to estimate $\left\{(\widehat{R}_{Kl}, \widehat{a}_{Kl}): l = 1,...,L_K \right\}$. These estimates completely determine the optimal list-based and cost constrained decision rule for the $K^\text{th}$ interval, $\widehat{d}_K$. To estimate treatment rules for decisions $k=1,...,K-1$, let $Q^Z_k(h_k, a_k; \widehat{d}\ )$ and $Q^Y_k(h_k, a_k; \widehat{d}\ )$ denote the Q-functions for interval $k$ assuming that the optimal list-based and cost constrained rule is followed in future intervals. The unconstrained optimal treatment decision at interval $k$ is $\widetilde{d}_k = \argmax_{a_k} Q^Z_k(h_k, a_k; \widehat{d}\ )$. To approximate this rule within the class of list-based and cost constrained treatment rules, we may apply the steps listed above replacing the $K^{\text{th}}$ stage Q-functions with their $k^\text{th}$ stage equivalents, and the cost constraint $\tau_K$ with $\sum_{j=k}^K \tau_j$.

\par The proposed algorithm allows estimation of an optimal list-based regime under a set of interval-specific cost constraints. This work requires two important modifications of the algorithm described by \cite{zhang2018interpretable} for identifying globally optimal list-based regimes. First, our procedure requires that the optimal values of $(R_{kl}, a_{kl})$ result in a rule which satisfies some pre-specified cost constraint. To achieve this, we must be able to estimate the expected cost accrued under each intermediate decision rule. This motivates our second departure from the original algorithm. Equation \ref{eqn:dtr-algorithm} can be viewed as the sum of expected outcomes in three exhaustive groups: (1) those satisfying the current clause but not past clauses ($\mathbb{I}(h_{ki} \in R_{kl}, h_{ki} \in \widehat{G}_{kl}) = 1$), (2) those satisfying neither the current clause nor past clauses ($\mathbb{I}(h_{ki} \not\in R_{kl}, h_{ki} \in \widehat{G}_{kl}) = 1$), and (3) those satisfying at least one previous clause ($\sum_{j=1}^{l-1} \mathbb{I}(h_{ki} \in R_{kj}, h_{ki} \not\in \widehat{G}_{kj}) = 1$). In order to estimate the overall expected outcomes, we must consider the expected outcome in each of these groups. Because \cite{zhang2018interpretable} are interested in finding an optimal list-based regime without a cost constraint, at each clause they need only find $(R_{kl}, a_{kl})$ that maximize effectiveness among unassigned units, i.e. those in groups (1) and (2). Because the outcomes for those in group (3) are fixed with respect to the choice of $(R_{kl}, a_{kl})$, these two methods will result in equivalent rules in cases where there is no cost constraint. This modification is necessary for ensuring the overall expected cost is below the cost constraint. A final important clarification concerns adjustments for censored individuals. While Q-functions for a given interval must be estimated using only those patients with observed outcomes, all patients who are uncensored at the beginning of a decision interval may be used to estimate optimal clauses within the decision list. This is true because the objective functions described in Equation \ref{eqn:dtr-algorithm} only require covariate histories, $H_k$, for each individual. This information is available for all units who have neither died nor been censored prior to interval $k$.

\subsection{Efficient optimization of treatment rules}
\par We propose an algorithm for identifying optimal choices of $R_{kl}$ and $a_{kl}$, at clause $l$ of decision rule $k$. Let $\{\widehat{R}_{kj}, \widehat{a}_{kj}: j=1,...,l-1 \}$ denote estimated optimal regions and treatments for clauses prior to $l$. Recall that regions $R_{kl}$ are defined in terms of a threshold, $\theta$, on a variable within $H_k$. Let $H_{kp}$, for $1 \le p \le \dim(H_k)$, denote a covariate contained within $H_k$. First, consider decision regions of the form $R_{kl} = \{h_{k}:h_{kp} \le \theta\}$ where patients with $h_k \in R_{kl}$ are given treatment $a_{kl}$. For $l < L_k$ the intermediate decision rule is given by:
\begin{itemize}
    \item[] \textbf{If} $h_{k} \in \widehat{R}_{k1}$ then $A_k = \widehat{a}_{k1}$;
    \item[] \textbf{else if} $h_{k} \in \widehat{R}_{k2}$ then $A_k = \widehat{a}_{k2}$;
    \item[] $\vdots$
    \item[] \textbf{else if} $h_{k} \in \widehat{R}_{k,l-1}$ then $A_k = \widehat{a}_{k,l-1}$;
    \item[] \textbf{else if} $h_{kp} \le \theta$ then $A_k = a_{kl}$;
    \item[] \textbf{else} $A_k = \widetilde{d}(h_k)$
\end{itemize}
For each unit, define:
\begin{align*}
    U_i^Z &= \mathbb{I}(h_i \in \widehat{G}_l)\widehat{Q}^Z_k(a_{kl}, h_i) + \sum_{j=1}^{l-1}\mathbb{I}(h_i \in \widehat{R}_j, h_i \not\in \widehat{G}_j)\widehat{Q}^Z_k(a_{kj}, h_i) \\
    V_i^Z &= \mathbb{I}(h_i \in \widehat{G}_l)\widehat{Q}^Z_k(\widetilde{d}(h_i), h_i) + \sum_{j=1}^{l-1}\mathbb{I}(h_i \in \widehat{R}_j, h_i \not\in \widehat{G}_j)\widehat{Q}^Z_k(a_{kj}, h_i)
\end{align*}
The values $U^Z_i$ and $V^Z_i$ represent expected effectiveness outcomes for patients included in and excluded from $R_{kl}$, respectively. Note that if a patient satisfies a previous clause, then $U^Z_i = V^Z_i$ so that the patient's expected outcome is independent of the choice of $\theta$ and $a_{kl}$. It can be seen that $\Psi^Z_{kl}(R_{kl}, a_{kl}) = \frac{1}{n} \sum_{i=1}^n \mathbb{I}(h_{kpi} \le \theta) U_i^Z + \mathbb{I}(h_{kpi} > \theta)V_i^Z$. If we define $U_i^Y$ and $V_i^Y$ similarly, it follows that $\Psi^Y_{kl}(R_{kl}, a_{kl}) = \frac{1}{n} \sum_{i=1}^n \mathbb{I}(h_{kpi} \le \theta) U_i^Y + \mathbb{I}(h_{kpi} > \theta)V_i^Y$. 
\par Although $\theta$ may take any real value, the unique values of $\Psi^Z_{kl}$ and $\Psi^Y_{kl}$ can be obtained by setting $\theta$ to the order statistics of $h_{kp}$. Let $\theta^*_1, ..., \theta^*_m$ denote the unique values of $H_{kp}$, where $m \le n$. It can be seen that $\Psi^Z$ and $\Psi^Y$ follow the recursive relationships:
\begin{align*}
    \Psi^Z_{kl}\big(\theta = \theta^*_j, a_{kl}) \big) &= \Psi^Z_{kl}\big(\theta = \theta^*_{j-1}, a_{kl} \big) + \sum_{i=1}^n \mathbb{I}(h_{kpi} = \theta^*_j )(U_i^Z - V_i^Z) \\
    \Psi^Y_{kl}\big(\theta = \theta^*_j, a_{kl}) \big) &= \Psi^Y_{kl}\big(\theta = \theta^*_{j-1}, a_{kl} \big) + \sum_{i=1}^n \mathbb{I}(h_{kpi} = \theta^*_j )(U_i^Y - V_i^Y).
\end{align*}
These relationships allow us to quickly enumerate all possible values of $\Psi^Z$ and $\Psi^Y$ for regions of the form $R_{kl} = \{h_{k}:h_{kp} \le \theta\}$ and assigned treatment $a_{kl}$. Analogous relationships can be defined for regions of the form $R_{kl} = \{h_{k}:h_{kp} > \theta\}$. Optimal choices for $R_{kl}$ and $a_{kl}$ can be obtained by iterating over each variable in $H_k$ and each possible treatment option, and identifying which choices lead to maximal values of $\Psi^Z(R_{kl}, a_{kl}) + \eta \sum_{i=1}^n \mathbb{I}(H_{ki} \in R_{kl})$ under the restriction $\Psi^Y_{R_{kl}, a_{kl}} < \sum_{j=k}^K \tau_k$.

\section{Simulation study}\label{sec:sim}

\refstepcounter{table}
\setcounter{table}{0}
\begin{table}[b]
    \label{tab:sim-res-numeric}
    \centering
         \footnotesize\begin{tabular}{lccccccccccc}
\toprule
& & &  \multicolumn{4}{c}{\underline{Linear Model}} & \multicolumn{4}{c}{\underline{SuperLearner}} \\ [0.1cm]
Corr. & $\tau$ & n & $\widehat{\text{Cost}}$ & $\widehat{\text{Surv.}}$ & MC. Cost & MC.  Surv. & & $\widehat{\text{Cost}}$ & $\widehat{\text{Surv.}}$ & MC. Cost & MC.  Surv. \\
\midrule
\midrule
Low & 26 & 500 & 25.73 & 1.46 & 25.25 & 1.42 & & 25.23 & 1.62 & 25.57 & 1.62\\
 &  & 1000 & 25.81 & 1.43 & 24.15 & 1.43 & & 25.36 & 1.65 & 25.80 & 1.70\\
 &  & 5000 & 25.98 & 1.39 & 23.70 & 1.42 & & 25.90 & 1.70 & 26.21 & 1.77\\
\addlinespace
 & 28 & 500 & 27.69 & 1.50 & 26.89 & 1.47 & & 27.78 & 1.65 & 27.11 & 1.67\\
 &  & 1000 & 27.79 & 1.46 & 25.70 & 1.46 & & 27.82 & 1.70 & 27.36 & 1.76\\
 &  & 5000 & 28.02 & 1.42 & 25.72 & 1.44 & & 27.87 & 1.76 & 28.05 & 1.84\\
\addlinespace
 & 30 & 500 & 29.71 & 1.54 & 28.92 & 1.50 & & 29.42 & 1.69 & 28.69 & 1.72\\
 &  & 1000 & 29.86 & 1.49 & 27.76 & 1.48 & & 29.79 & 1.74 & 29.24 & 1.80\\
 &  & 5000 & 29.99 & 1.45 & 27.52 & 1.47 & & 29.83 & 1.80 & 29.89 & 1.88\\
\addlinespace
Med. & 26 & 500 & 25.57 & 1.44 & 25.59 & 1.41 & & 25.76 & 1.56 & 25.24 & 1.56\\
 &  & 1000 & 25.71 & 1.39 & 24.26 & 1.40 & & 25.84 & 1.59 & 25.81 & 1.64\\
 &  & 5000 & 25.97 & 1.36 & 23.55 & 1.40 & & 25.94 & 1.63 & 26.40 & 1.69\\
\addlinespace
 & 28 & 500 & 27.65 & 1.46 & 27.02 & 1.42 & & 27.26 & 1.60 & 27.57 & 1.62\\
 &  & 1000 & 27.78 & 1.42 & 25.87 & 1.42 & & 27.78 & 1.63 & 27.44 & 1.68\\
 &  & 5000 & 27.99 & 1.38 & 25.32 & 1.41 & & 27.91 & 1.68 & 28.26 & 1.75\\
\addlinespace
 & 30 & 500 & 29.71 & 1.48 & 28.70 & 1.45 & & 29.53 & 1.63 & 28.89 & 1.65\\
 &  & 1000 & 29.74 & 1.44 & 27.20 & 1.45 & & 29.79 & 1.68 & 29.42 & 1.74\\
 &  & 5000 & 30.00 & 1.41 & 27.16 & 1.43 & & 29.89 & 1.74 & 30.11 & 1.80\\
\addlinespace
High & 26 & 500 & 25.56 & 1.41 & 25.69 & 1.37 & & 25.77 & 1.52 & 25.66 & 1.51\\
 &  & 1000 & 25.65 & 1.37 & 24.52 & 1.38 & & 25.64 & 1.53 & 25.75 & 1.56\\
 &  & 5000 & 25.87 & 1.33 & 23.55 & 1.38 & & 25.95 & 1.56 & 26.51 & 1.62\\
\addlinespace
 & 28 & 500 & 27.44 & 1.43 & 27.23 & 1.40 & & 27.72 & 1.54 & 27.40 & 1.55\\
 &  & 1000 & 27.66 & 1.39 & 25.91 & 1.40 & & 27.70 & 1.58 & 27.66 & 1.62\\
 &  & 5000 & 27.95 & 1.36 & 25.18 & 1.39 & & 27.95 & 1.62 & 28.44 & 1.68\\
\addlinespace
 & 30 & 500 & 29.56 & 1.45 & 29.12 & 1.42 & & 29.71 & 1.58 & 29.09 & 1.59\\
 &  & 1000 & 29.63 & 1.41 & 27.21 & 1.42 & & 29.74 & 1.62 & 29.57 & 1.66\\
 &  & 5000 & 29.99 & 1.38 & 26.73 & 1.41 & & 29.92 & 1.66 & 30.40 & 1.72\\
\bottomrule
\end{tabular}
\caption{Mean value of Monte Carlo estimates of the mean survival (MC. Surv.) and cost (MC. Cost), and mean estimated survival ($\widehat{\text{Surv.}}$) and cost ($\widehat{\text{Cost}}$) arising from averaging over the Q-functions. Results are provided for simulations with 30\% censoring and with low, medium, and high levels of correlation and sample sizes (n).}
\end{table}

\par We perform a simulation study to (1) demonstrate that our approach is able to fit valid cost constrained regimes in realistic settings and (2) illustrate the utility of flexible ensemble learning approaches for improving regime performance. We simulate data with $K = 3$ decision points and $n = 500$, $1000$, and $5000$ sample units. At each time point, we simulate a vector of independent standard normal confounding variables, $W_{k}$; a binary treatment decision, $A_k$; an indicator of whether a unit is censored during an interval, $C_k$; an indicator of whether a unit has survived to the end of the interval, $S_k$; and cost accrued within the interval, $Y_k$. Simulations are run at various sample sizes and levels of correlation between survival and cost outcomes. Details for how data are simulated can be found in the appendix. We compare two approaches for modeling the Q-functions: (1) linear models, and (2) SuperLearner. SuperLearner is an ensemble learner, combining predictions from a set of candidate learners to minimize cross-validated risk \cite{van2011targeted}. The candidate learners within our SuperLearner are random forests, neural nets, elastic net, and generalized linear models.
\par For each simulation setting, we generate $500$ datasets and estimate an optimal cost constrained regimes under overall cost constraints of $\tau = 26$, $28$, and $30$. After regimes are identified, we simulate $10^6$ new units, treat them according to the estimated treatment rules, and record the mean survival and cost among these units. This procedure results in two estimates of the mean cost and survival under each fit regime; one obtained by averaging over the estimated Q-functions for the initial interval, and another based on Monte Carlo simulation. The Monte Carlo approach allows us to determine how each regime performs when applied to new data and approximates the true cost and survival under that regime. If the Q-functions are appropriately modelled, then estimates of the expected cost and survival for a fit regime should be similar to Monte Carlo approximations of the truth.

\par Table \ref{tab:sim-res-numeric} provides the mean Q-function-based estimate of cost and survival as well as the mean Monte Carlo estimated cost and survival across regimes fit under varying cost constraints and levels of correlation between cost and effectiveness outcomes. Results in this table are from settings with approximately $30\%$ censoring over the course of the three treatment intervals. Results at $0\%$ and $60\%$ censoring are provided in the appendix. In nearly all settings, the mean Monte Carlo estimated cost fall below the specified cost threshold. Only in scenarios where SuperLearner was used with large sample datasets, was the cost constraint not satisfied. In these settings, the mean Q-function based estimate of the total cost underestimated the true Monte-Carlo estimated cost (e.g. true vs. estimated cost of $30.40$ vs. $29.92$, when $n=5000$, $\tau = 30$, and under high correlation). Due to misspecification, regimes identified using linear models consistently overestimated the costs observed among new units. Because total regime costs are overestimated when using linear models, some regimes which would truly satisfy the cost constraint may be excluded from consideration. This results in more conservative treatment strategies and decreased survival when compared to SuperLearner-based regimes. In the simulation setting with $n=1000$, $\tau=28$, and medium correlation between outcomes, both Linear-model and SuperLearner-based regimes have a mean estimated treatment cost of $27.78$. However, because the SuperLearner-based models more accurately estimate the true regime cost ($25.87$ and $27.44$ for linear-model and SuperLearner-based regimes, respectively), resources are better utilized and units acheive greater survival ($1.42$ vs. $1.68$). Additionally, because SuperLearner may improve our ability to identify patients that are likely to benefit from treatment, Monte Carlo estimates of survival in SuperLearner-based regimes exceed those of Linear-model-based regimes even when costs are similar (e.g. Linear model vs. SuperLearner-based survival of $1.37$ vs. $1.51$ and costs of $25.69$ vs. $25.66$ when $n=500$, $\tau = 26$, and under high correlation). 



\section{The cost-effectiveness of adjuvant therapies for endometrial cancer}\label{sec:data_example}

\par The standard treatment for patients with early-stage endometrial cancer is complete hysterectomy. Throughout the post-surgical period, patients may receive adjuvant radiation or chemotherapy to decrease the risk of recurrence \citep{latif2014adjuvant}. The decision to provide adjuvant therapy will depend on individual patient characteristics. In particular, younger patients or those with low-grade histology have low risk of recurrence and may not benefit from adjuvant therapies \citep{creutzberg2000surgery}. Additionally, the best treatment decision for a given patient will change over time. \cite{hogberg2010sequential} found that certain high risk patients who initially receive adjuvant chemotherapy may exhibit greater survival if switched to radiation therapy. However, the optimal sequencing of adjuvant therapies remains a subject of controversy \citep{van2021adjuvant}. In this analysis, we use data from the linked SEER-Medicare database to identify an optimally cost-effective regime for assigning adjuvant treatments to endometrial cancer patients. In doing so, we aim to explore how different treatment strategies effect patient survival and costs. Patients in this database were diagnosed with endometrioid histology cancer between $2000$ and $2011$, with follow-up until $2013$. 
\par Each patient in the database is followed for $24$ months. We divide this period into four intervals (months 1-6, 7-12, 13-18, and 19-24), wherein treatment is assigned within the first month of each interval. At the beginning of each interval, patients are assigned one of three treatments: (1) adjuvant radiation therapy (RT), (2) adjuvant chemotherapy (CT), or (3) monitoring alone (neither radiation nor chemotherapy), hereafter referred to as control. The data contain followup information on $13722$ patients, $714$ of whom have censored outcomes. Of the total study population, $27$ patients were excluded because they were treated with both adjuvant RT and CT. At baseline, data are available on patient's age, race, and cancer stage and grade. Additionally, Charlson comorbidity indices and the number of hospitalizations are recorded at every month of follow-up. The mean age at diagnosis is $73.72$ (SD = $6.58$), and most patients have stage I cancer ($94.03\%$). Charlson comorbidity indices are between zero and five, with $54.39\%$ of patients having an index of zero. The 24-month-restricted mean survival among patients is $22.86$ months and the mean accumulated cost over the study period is $33830.89$.
\par Due to sensitivity from random seeds when using SuperLearner, we fit Q-functions using ordinary least squares regression. Our outcomes of interest were total cost over the 24 month period and restricted mean survival. Covariates used as predictors include: age at diagnosis, race, cancer stage, cancer grade, number of hospitalizations over the previous six months, maximum Charlson comorbidity index over the previous six months, total cost accrued over the previous six months, and previous treatment assignment. Treatment rules for each interval contain a maximum of four clauses and assign adjuvant therapies based upon a patient's age at diagnosis, Charlson comorbidity index, number of hospitalizations over the past 6 months, previous treatments, and their cancer stage and grade.
\par Because medical costs do not accrue at the same rate throughout the post-surgical period, we allow the interval specific cost constraints for each candidate regime to change with time. We parameterize our candidate regimes using $s \in [0, 1]$. The vector of interval-specific cost constraints corresponding to $s$ is given by $\bm{\tau}_{s} = (1-s)(18500, 5500, 5500, 5000)^T + s(18500, 5500, 9000, 8000)^T$. For $s=0$, the interval-specific constraints approximate the pattern of cost-accrual observed among patients within the database, while for $s=1$, the constraints mirror the cost-accrual pattern under an optimal regime fit without resource constraints. Candidate regimes are fit under the vector of cost constraints corresponding to $s=0, 0.2, 0.4, ..., 1
$.

\par To identify an optimally cost-effective regime, we sequentially compare each candidate regime using the ICER. Given two candidate regimes, $d_1$ and $d_2$, ICER is a comparative measure defined as the ratio of the difference in expected cost to the difference in expected effectiveness between the two regimes:
\begin{equation*}
    \text{ICER}(d_2, d_1) = \frac{\mathbb{E}[Y^{d_2}] - \mathbb{E}[Y^{d_1}]}{\mathbb{E}[Z^{d_2}] - \mathbb{E}[Z^{d_1}]}.
\end{equation*}
ICER  can be interpreted as the cost per unit change in effectiveness obtained by switching from treatment regime $d_1$ to $d_2$. Let $\lambda$ denote a pre-selected willingness-to-pay (WTP) parameter. The WTP represents the maximum cost a payer is willing to incur for a unit change in effectiveness. Adopting regime $d_2$ over the comparator regime, $d_1$, is considered cost-effective if ICER is less than the chosen WTP. Given a set of regimes corresponding to different cost-constraints, $\widehat{d}(\tau_1)$, ..., $\widehat{d}(\tau_J)$, we provide an algorithm that identifies the optimally cost-effective regime by sequentially comparing each candidate regime to the best existing alternative. 
\begin{enumerate}
    \item[Step 1.] Define the current most cost-effective regime: $\widehat{d}_{\text{CE}} \coloneqq \widehat{d}(\tau_1)$ and set $j = 2$
    \item[Step 2.] If $\left(\text{ICER}(\widehat{d}(\tau_j), \widehat{d}_{\text{CE}}) < \lambda\right)$, then $\widehat{d}_{\text{CE}} \coloneqq \widehat{d}(\tau_j) $, otherwise do not update $\widehat{d}_{\text{CE}}$.
    \item[Step 3.] If $j < J$, set $j = j+1$ and repeat from Step 2, otherwise $\widehat{d}_{\text{CE}}$ is the optimally cost-effective regime. 
\end{enumerate}
\par We perform this cost-effectiveness analysis to select an optimally cost-effective regime from the set of regimes with cost-constraints given by $\tau_s$ for $s = 0, 0.2, 0.4, ..., 1$. Results of this analysis under a WTP of $\$4{,}166$/month or equivalently, $\$50{,}000$/year are provided in Table \ref{tab:endometrial_ce_table}. Because none of the regimes have ICER less than this WTP when compared with Regime I, Regime I (the least expensive regime) is the optimally cost-effective list-based treatment regime. The treatment rules for this regime are provided within the appendix. Under this regime, patients are expected to accrue approximately $\$34{,}013$ in medical costs and to survive for $22.837$ months within the two years following hysterectomy. Estimated restricted mean survival and cost under this regime are similar to those observed among patients in our database. 
\refstepcounter{table}
\setcounter{table}{1}
\begin{table}[t]
    \label{tab:endometrial_ce_table}
    \centering
\begin{tabular}{ccccc}
\toprule
Regime & Cost & Survival & ICER & Comparator\\
\midrule
\midrule
\textbf{I} & \textbf{34012.72} & \textbf{22.837} & \textbf{NA} & \textbf{NA}\\
II & 35727.89 & 22.861 & 71467.09 & I\\
III & 37195.06 & 22.944 & 29591.78 & I\\
IV & 38018.09 & 22.966 & 30990.00 & I\\
V & 39959.90 & 22.972 & 43890.00 & I\\
\bottomrule
\end{tabular}
\caption{Estimated mean survival (months) and cost (USD\$) for patients treated according to each candidate regime. Incremental cost-effectiveness ratios comparing subsequent candidate regimes. Results for optimally cost-effective treatment regime at WTP of $\$4{,}166$/month bolded.}
\end{table}

\refstepcounter{figure}
\setcounter{figure}{0}
\begin{figure}[b]
\centering

\subfloat[]{\includegraphics[width=6.5cm]{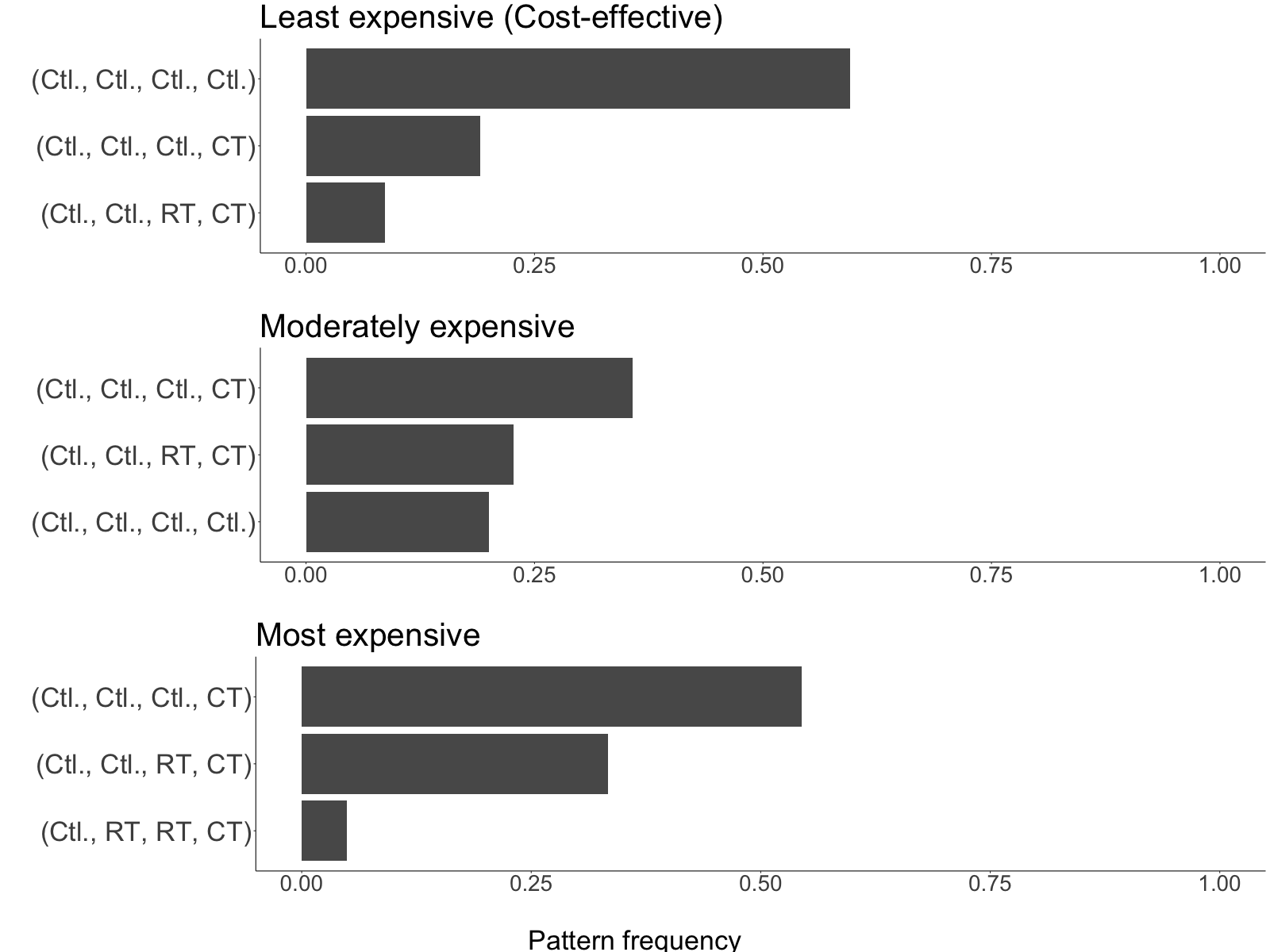}}\quad  \label{fig:sub1}
\subfloat[]{\includegraphics[width=6.5cm]{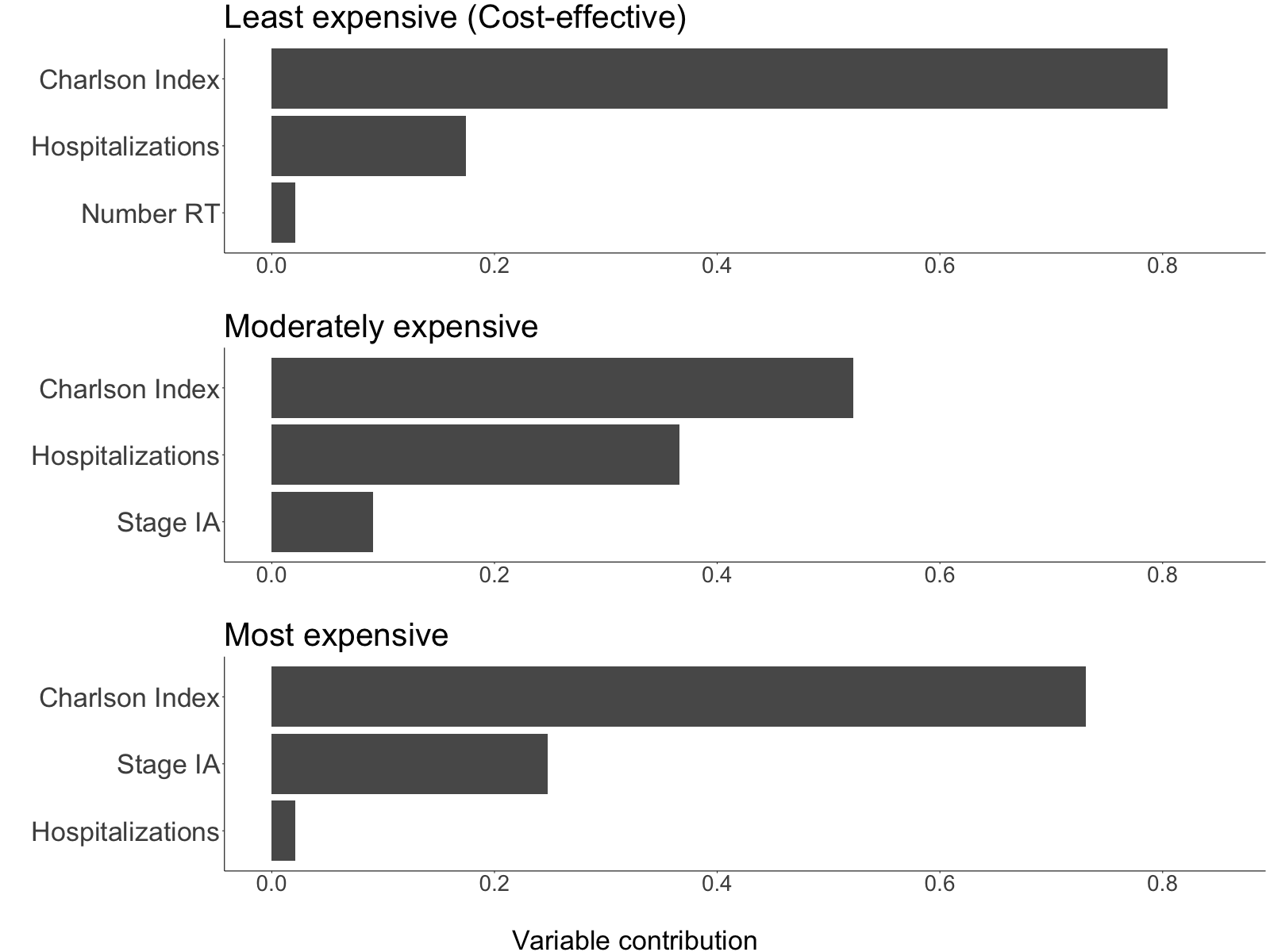}}  \label{fig:sub2}

\caption{Most common treatment patterns (a) and variables (b) for the optimally cost-effective, a moderately expensive, and most expensive treatment regimes. Estimated by determining which treatments the observed units would be assigned to based on their covariate history.}
\end{figure}

\par We may characterize the differences between the optimal regime and other candidate regimes is by comparing treatment patterns under each regime. Figure \ref{fig:sub1} provides the most frequent treatment patterns under the optimal, a moderately expensive (regime III), and the most expensive regimes. Under the optimal regime the three most common treatment patterns are: (1) Neither treatment for 2.0 years (60\%), (2) Neither treatment for the first 1.5 years and chemotherapy for 0.5 years (19\%), and (3) neither adjuvant therapy for 1.0 years, radiation for 0.5 years, and chemotherapy for the final 0.5 years (9\%). The most common treatment patterns under regime IV were the same, but in a different order of importance. Both of these differ from the most expensive regime, under which less than 5\% of patients receive neither adjuvant therapy for the duration of study. In practice, treatment with radiation or chemotherapy after the initial monitoring period often indicates cancer recurrence. Because our data do not contain information on recurrence, other covariates must be used as proxies to determine whether recurrence has occurred and additional treatment is required. The most common treatment patterns in the more expensive regimes entail treatment after a period of monitoring, suggesting that additional costs are incurred due to more aggressive predictions of recurrence. We may also compare which variables are most frequently selected to assign treatment between regimes using Figure \ref{fig:sub2}. For all three considered regimes, treatment decisions were most commonly based on a patient's Charlson comorbidity index. Number of hospitalizations over the past six months, the presence of stage IA cancer, and the number of RT sessions they had undergone over the previous interval also helped differentiate which patients would most benefit from each treatment.

\section{Discussion}\label{sec:discussion}

\par In this paper, we present a two-step procedure for identifying an optimally cost-effective dynamic treatment regime with interpretable, list-based decision rules. In the first step, we use a novel Q-learning and policy search-based approach to estimate optimal list-based regimes that maximize treatment effectiveness under a predefined constraint on treatment costs. Through simulations we show the validity of our cost constrained regimes and illustrate how flexible ensemble learners can improve regime performance. The second step consists of a cost-effectiveness analysis that can select an optimally cost-effective regime from a set of candidate regimes characterized by variable cost constraints. This procedure identifies the most cost-effective treatment strategy by iteratively comparing each candidate regime with the best existing alternative. 

\par While our method works for general measures of effectiveness, in our simulations and data example we focus on restricted mean survival. This is a traditional measure of effectiveness within cost-effectiveness studies \citep{li2018doubly, spieker2019net}. Recently, \cite{linn2017interactive} developed methods for optimizing specified quantiles of a distribution. Future extensions may explore how this methodology could be used to allow for maximization of median survival time while placing a constraint on overall cost. In contrast with previous work on identifying cost-effective regimes which have maximized effectiveness while penalizing costs \citep{lakkaraju2017learning, xu2020estimating}, our proposed cost-effectiveness analysis is based upon iterative comparisons of candidate regimes. This allows us to better characterize how decisions change as the allowable cost increases and avoids complications arising from non-iterative cost-effectiveness analyses \citep{cohen2008interpreting}. For example, suppose two or more experimental treatments are considered cost-effective when compared with the standard of care. Additional iterative comparisons between the experimental treatments are necessary to determine which treatment is preferred under a pre-specified budgetary constraint. 

\par An important consideration in the design of cost-effectiveness analyses is in the selection of cost constraints. In general, if too few cost constraints (and thus, too few candidate regimes) are selected, then we may be unable to approximate a true optimally cost-effective regime. If cost constraints are too similar, then ICER may become unstable due to similar estimated effectiveness. This consideration is further complicated by difficulties in quantifying uncertainty. Constrained estimation results in nonstandard asymptotic theory, and methods for performing inference have not yet been explored \citep{laber2018identifying}. Additionally, the modeling choice for Q-functions may influence estimated treatment rules. Flexible methods like SuperLearner may improve regime performance, but are also subject to variability across random seeds. For estimates of treatment effects, \cite{benkeser2020practical} propose combining estimates arising from different seeds. However, work is required to determine the best procedure for combining decision-list based rules. 

\bibliographystyle{imsart-nameyear}
\bibliography{DTRCE.bib}

\clearpage
\begin{appendix}
\section*{Simulation Setup}
Baseline data are simulated as: 

\footnotesize\begin{itemize}
    \item $W_{1} \sim N_2(0, 1)$
    \item $P(C_1) = p_c$
    \item $P(A_1|W_1) = \text{expit}(-0.8 + \beta_1^T W_1)$
    \item $P(S_1|A_1, W_1) = \text{expit}\bigg\{-0.25 + \alpha_1^T W_1 + 2A_1 + \alpha_{2}^T (A_1 W_1) + 1.2\Phi(\alpha_3^T A_1 W_1) - \\ 3 A_1 \mathbb{I}\left[\min(W_1) < -1\right]\bigg\}$
    \item $\log(Y_1) \sim N\{1 + \eta_0 + \eta_1^T W_1 + 0.5A_1 + 1.5 \sum_{j=1}^2 \mathbb{I}[W_{1j} > 1.5]*A_1 + \eta_2^T A_1W_1 + \\ \zeta P(S_1|A_1, W_1),\ 0.075\}$
\end{itemize}
\normalsize and data for intervals $k=2,3$ are simulated as,
\footnotesize\begin{itemize}
    \item $W_{k} \sim N_2(0, 1)$
    \item $P(C_k) = p_c$
    \item $P(A_k|W_k, A_{k-1}, W_{k-1}) = \text{expit}\left\{-0.8 + \beta_{1F}^T W_k + \beta_{2F}^T W_{k-1} + 0.5 A_{k-1}\right\}$
    \item $P(S_k|\overline{A}_k, \overline{W}_k) =  \text{expit}\bigg\{-0.25 + \alpha_{1F}^T W_k + 2A_k + \alpha_{2F}^T (A_k W_{k}) + 1.2\Phi(\alpha_{3F}^T A_k W_k) - \\  3 A_1 \mathbb{I}\left[\min(W_k) < -1\right]+ \alpha_{4F}^T W_{k-1}\bigg\}$
    \item $\log(Y_k) \sim N\{1 + \eta_{0F} + \eta_{1F}^T W_k + 0.5A_k + 1.5 \sum_{j=1}^2 \mathbb{I}[W_{kj} > 1]*A_k + \eta_{2F}^T A_kW_k + \\ \zeta P(S_k|A_k, W_k, A_{k-1}, W_{k-1}),\ 0.1 \}$
\end{itemize}
\normalsize We define $\beta_1 = \{0.4, 0.4\}$, $\alpha_1 = \{-0.4, -0.4 \}$, $\alpha_2 = \{ 0.5, 0.125 \}$, $\alpha_3 = \{ 1, 0.25 \}$, $\eta_1 = \{0.2, 0.2\}$, and $\eta_2 = \{ 0.3, 0.1 \}$. For follow-up intervals the corresponding parameter values are $\beta_{1F} = \{0.3, 0.3 \}$, $\beta_{2F} = \{0.1, 0.1\}$, $\alpha_{1F} = \{-0.3, -0.3 \}$, $\alpha_{2F} = \{0.6, 0.125 \}$, $\alpha_{3F} = \{ 1, 0.25 \}$, $\alpha_{4F} = \{-0.1, -0.1 \}$ $\eta_{1F} = \{0.2, 0.2 \}$, and $\eta_{2F} = \{ 0.3, 0.1 \}$. The parameter $\zeta$, which is the same for baseline and follow-up data, controls the level of correlation between the cost and effectiveness outcomes. We consider simulation settings with low, medium, and high levels of correlation between outcomes, corresponding to $\zeta = 0.5, 1$, and $1.5$ respectively. In addition, we consider simulations with 5\%, 20\%, and 50\% censoring. These correspond to $p_c = 1/60$, $2/30$, and $1/6$, respectively. To ensure costs are similar across simulation settings we include the offset term $\alpha_0$ which is equal to $0.62, 0.31$ and $0.0$ in low, medium, and high correlation settings.

\clearpage

\section{Simulation results at varied levels of censoring}

\begin{table}[b]
    \label{tab:sim-res-numeric2}
    \centering
         \footnotesize\begin{tabular}{lccccccccccc}
\toprule
& & &  \multicolumn{4}{c}{\underline{Linear Model}} & \multicolumn{4}{c}{\underline{SuperLearner}} \\ [0.1cm]
Corr. & $\tau$ & n & $\widehat{\text{Cost}}$ & $\widehat{\text{Surv.}}$ & MC. Cost & MC.  Surv. & & $\widehat{\text{Cost}}$ & $\widehat{\text{Surv.}}$ & MC. Cost & MC.  Surv. \\
\midrule
\midrule
Low & 26 & 500 & 25.72 & 1.48 & 25.60 & 1.43 & & 24.99 & 1.61 & 26.00 & 1.60\\
 &  & 1000 & 25.77 & 1.44 & 24.21 & 1.43 & & 25.86 & 1.64 & 25.64 & 1.68\\
 &  & 5000 & 26.01 & 1.39 & 23.86 & 1.42 & & 25.89 & 1.69 & 26.17 & 1.77\\
\addlinespace
 & 28 & 500 & 27.80 & 1.52 & 28.61 & 1.48 & & 27.63 & 1.64 & 27.49 & 1.64\\
 &  & 1000 & 27.78 & 1.47 & 26.25 & 1.47 & & 27.81 & 1.69 & 27.26 & 1.73\\
 &  & 5000 & 27.98 & 1.42 & 25.68 & 1.45 & & 27.87 & 1.75 & 28.02 & 1.84\\
\addlinespace
 & 30 & 500 & 29.73 & 1.56 & 29.46 & 1.49 & & 29.65 & 1.67 & 28.56 & 1.67\\
 &  & 1000 & 29.81 & 1.50 & 27.97 & 1.49 & & 29.73 & 1.71 & 28.76 & 1.77\\
 &  & 5000 & 29.97 & 1.45 & 27.51 & 1.47 & & 29.82 & 1.80 & 29.87 & 1.89\\
\addlinespace
Med. & 26 & 500 & 25.60 & 1.46 & 25.79 & 1.40 & & 25.46 & 1.56 & 25.48 & 1.54\\
 &  & 1000 & 25.65 & 1.40 & 24.37 & 1.40 & & 25.79 & 1.57 & 25.49 & 1.61\\
 &  & 5000 & 25.99 & 1.36 & 23.64 & 1.40 & & 25.94 & 1.62 & 26.35 & 1.69\\
\addlinespace
 & 28 & 500 & 27.63 & 1.48 & 28.09 & 1.44 & & 27.60 & 1.59 & 27.34 & 1.58\\
 &  & 1000 & 27.65 & 1.44 & 25.81 & 1.43 & & 27.78 & 1.63 & 27.31 & 1.66\\
 &  & 5000 & 27.97 & 1.38 & 25.33 & 1.41 & & 27.92 & 1.68 & 28.32 & 1.75\\
\addlinespace
 & 30 & 500 & 29.66 & 1.52 & 29.73 & 1.45 & & 27.21 & 1.64 & 29.67 & 1.62\\
 &  & 1000 & 29.80 & 1.46 & 28.18 & 1.45 & & 29.76 & 1.67 & 29.20 & 1.71\\
 &  & 5000 & 29.97 & 1.41 & 27.20 & 1.43 & & 29.89 & 1.73 & 30.06 & 1.80\\
\addlinespace
High & 26 & 500 & 25.61 & 1.42 & 26.28 & 1.38 & & 25.19 & 1.54 & 26.55 & 1.50\\
 &  & 1000 & 25.67 & 1.38 & 24.58 & 1.37 & & 25.73 & 1.53 & 26.15 & 1.56\\
 &  & 5000 & 25.90 & 1.34 & 23.71 & 1.38 & & 25.96 & 1.56 & 26.54 & 1.63\\
\addlinespace
 & 28 & 500 & 27.56 & 1.46 & 28.36 & 1.41 & & 27.58 & 1.56 & 27.61 & 1.52\\
 &  & 1000 & 27.70 & 1.40 & 26.37 & 1.39 & & 27.68 & 1.57 & 27.47 & 1.60\\
 &  & 5000 & 27.96 & 1.35 & 25.13 & 1.39 & & 27.93 & 1.61 & 28.45 & 1.68\\
\addlinespace
 & 30 & 500 & 29.60 & 1.47 & 30.41 & 1.42 & & 29.08 & 1.58 & 29.48 & 1.57\\
 &  & 1000 & 29.76 & 1.42 & 27.86 & 1.42 & & 29.78 & 1.61 & 29.40 & 1.65\\
 &  & 5000 & 29.94 & 1.38 & 26.87 & 1.40 & & 29.93 & 1.65 & 30.31 & 1.72\\
\bottomrule
\end{tabular}
\caption{Mean value of Monte Carlo estimates of the mean survival (MC. Surv.) and cost (MC. Cost), and mean estimated survival ($\widehat{\text{Surv.}}$) and cost ($\widehat{\text{Cost}}$) arising from averaging over the Q-functions. Results are provided for simulations with 60\% censoring and with low, medium, and high levels of correlation and sample sizes (n).}
\end{table}

\begin{table}[b]
    \label{tab:sim-res-numeric3}
    \centering
         \footnotesize\begin{tabular}{lccccccccccc}
\toprule
& & &  \multicolumn{4}{c}{\underline{Linear Model}} & \multicolumn{4}{c}{\underline{SuperLearner}} \\ [0.1cm]
Corr. & $\tau$ & n & $\widehat{\text{Cost}}$ & $\widehat{\text{Surv.}}$ & MC. Cost & MC.  Surv. & & $\widehat{\text{Cost}}$ & $\widehat{\text{Surv.}}$ & MC. Cost & MC.  Surv. \\
\midrule
\midrule
Low & 26 & 500 & 25.72 & 1.45 & 25.14 & 1.44 & & 25.77 & 1.61 & 25.14 & 1.65\\
 &  & 1000 & 25.84 & 1.42 & 23.80 & 1.43 & & 25.83 & 1.66 & 25.76 & 1.72\\
 &  & 5000 & 26.03 & 1.39 & 23.73 & 1.42 & & 25.90 & 1.70 & 26.16 & 1.77\\
\addlinespace
 & 28 & 500 & 27.79 & 1.48 & 26.65 & 1.46 & & 27.79 & 1.66 & 26.82 & 1.70\\
 &  & 1000 & 27.85 & 1.45 & 25.90 & 1.46 & & 27.82 & 1.70 & 27.51 & 1.78\\
 &  & 5000 & 28.00 & 1.42 & 25.56 & 1.44 & & 27.88 & 1.76 & 28.02 & 1.84\\
\addlinespace
 & 30 & 500 & 29.72 & 1.52 & 28.04 & 1.49 & & 29.73 & 1.70 & 28.68 & 1.74\\
 &  & 1000 & 29.81 & 1.48 & 27.50 & 1.49 & & 29.79 & 1.75 & 29.35 & 1.81\\
 &  & 5000 & 29.98 & 1.45 & 27.70 & 1.46 & & 29.84 & 1.81 & 29.83 & 1.88\\
\addlinespace
Med. & 26 & 500 & 25.55 & 1.42 & 24.70 & 1.40 & & 25.56 & 1.58 & 25.65 & 1.58\\
 &  & 1000 & 25.75 & 1.38 & 24.03 & 1.39 & & 25.82 & 1.59 & 26.00 & 1.65\\
 &  & 5000 & 26.04 & 1.36 & 23.61 & 1.40 & & 25.93 & 1.63 & 26.41 & 1.70\\
\addlinespace
 & 28 & 500 & 27.67 & 1.45 & 27.00 & 1.43 & & 27.71 & 1.61 & 27.27 & 1.63\\
 &  & 1000 & 27.73 & 1.42 & 25.80 & 1.43 & & 27.84 & 1.64 & 27.69 & 1.70\\
 &  & 5000 & 27.99 & 1.38 & 25.30 & 1.42 & & 27.92 & 1.68 & 28.25 & 1.75\\
\addlinespace
 & 30 & 500 & 29.67 & 1.48 & 28.35 & 1.46 & & 29.67 & 1.65 & 28.76 & 1.68\\
 &  & 1000 & 29.74 & 1.44 & 27.39 & 1.45 & & 29.76 & 1.69 & 29.54 & 1.75\\
 &  & 5000 & 29.97 & 1.41 & 27.00 & 1.43 & & 29.90 & 1.74 & 30.11 & 1.80\\
\addlinespace
High & 26 & 500 & 25.49 & 1.39 & 24.98 & 1.37 & & 25.76 & 1.51 & 25.60 & 1.53\\
 &  & 1000 & 25.71 & 1.36 & 24.29 & 1.37 & & 25.77 & 1.54 & 26.25 & 1.59\\
 &  & 5000 & 25.98 & 1.33 & 23.44 & 1.38 & & 25.96 & 1.56 & 26.49 & 1.62\\
\addlinespace
 & 28 & 500 & 27.58 & 1.42 & 27.10 & 1.39 & & 27.72 & 1.56 & 27.09 & 1.57\\
 &  & 1000 & 27.72 & 1.39 & 25.81 & 1.39 & & 27.80 & 1.59 & 27.91 & 1.63\\
 &  & 5000 & 27.97 & 1.35 & 24.96 & 1.40 & & 27.95 & 1.61 & 28.50 & 1.68\\
\addlinespace
 & 30 & 500 & 29.54 & 1.44 & 28.25 & 1.41 & & 29.65 & 1.59 & 28.95 & 1.61\\
 &  & 1000 & 29.72 & 1.41 & 27.50 & 1.42 & & 29.74 & 1.62 & 29.77 & 1.68\\
 &  & 5000 & 29.98 & 1.38 & 27.01 & 1.40 & & 29.93 & 1.66 & 30.33 & 1.73\\
\bottomrule
\end{tabular}
\caption{Mean value of Monte Carlo estimates of the mean survival (MC. Surv.) and cost (MC. Cost), and mean estimated survival ($\widehat{\text{Surv.}}$) and cost ($\widehat{\text{Cost}}$) arising from averaging over the Q-functions. Results are provided for simulations with 0\% censoring and with low, medium, and high levels of correlation and sample sizes (n).}
\end{table}

\clearpage

\section{Optimally Cost-effective treatment regime for adjuvant treatment of endometrial cancer}

The optimally cost-effective treatment regime (Regime I) for the adjuvant treatment of endometrial cancer is given by:
\begin{enumerate}
    \item Interval 1
    \begin{itemize}[noitemsep, topsep = 1pt, labelwidth = 1pt]
    \item[{ }] \textbf{All} receive $A_1 = 0$;
\end{itemize} 
\smallskip
\item Interval 2
    \begin{itemize}[noitemsep, topsep = 1pt, labelwidth = 1pt]
    \item[{ }] \textbf{If} $\textsc{Charlson Index} \le 2$  then $A_2 = 0$;
    \item[{ }] \textbf{else if} $\textsc{Number RT over past 6 mo.} = 0$ then $A_2 = 1$;
    \item[{ }] \textbf{else if} $\textsc{Number Hospitalizations} \le 1$ then $A_2 = 0$
    \item[{}] \textbf{else} $A_2 = 1$
\end{itemize}
\smallskip
\item Interval 3
    \begin{itemize}[noitemsep, topsep = 1pt, labelwidth = 1pt]
    \item[{ }] \textbf{If} $\textsc{Charlson Index} \le 1$  then $A_3 = 0$;
    \item[{ }] \textbf{else if} $\textsc{Number Hospitalizations} = 0$ then $A_3 = 1$
    \item[{ }] \textbf{else if} $\textsc{Charlson Index} = 2$ then $A_3 = 0$
    \item[{}] \textbf{else} $A_3 = 1$
\end{itemize}
\smallskip
\item Interval 4
    \begin{itemize}[noitemsep, topsep = 1pt, labelwidth = 1pt]
    \item[{ }] \textbf{If} $\textsc{Charlson Index} = 0$  then $A_4 = 0$;
    \item[{ }] \textbf{else if} $\textsc{Number Hospitalizations} = 0$ then $A_4 = 2$
    \item[{}] \textbf{else} $A_4 = 0$
\end{itemize}
\end{enumerate}

\end{appendix}

\end{document}